\documentstyle[prl,eqsecnum,aps]{revtex}
\twocolumn

\begin{document}
\author{J. Q. Shen $^{1,}$$^{2}$ \footnote{E-mail address: jqshen@coer.zju.edu.cn}}
\address{$^{1}$ Centre for Optical
and Electromagnetic Research, State Key Laboratory of Modern
Optical Instrumentation, \\Zhejiang University,
Hangzhou Yuquan 310027, P. R. China\\
$^{2}$ Zhejiang Institute of Modern Physics and Department of
Physics, Zhejiang University, Hangzhou 310027, P. R. China}
\date{\today }
\title{Comment on ``New Experimental Limit on the Photon Rest Mass with a Rotating Torsion Balance''}
\maketitle
   During the last three decades, photon rest mass
problem captured special attention of many investigators who have
reported several experimental upper limits on the photon mass by
using various
methods\cite{Feinberg,Williams,Barrow,Chernikov,Fischbach,Lakes}.
More recently, Luo {\it et al.} obtained the most new upper limit
on photon rest mass of $1.2\times 10^{-54}$ Kg by means of {\it
rotating torsion balance} experiment\cite{Luo}. Although Luo {\it
et al.}'s experiment is excellent, here we have a supplement to
their result, since our evaluation shows that the {\it photon
effective rest mass} due to the {\it self-induced charge
currents}\cite{Ho} in the environmental dilute plasma ({\it e.g.},
the muon component and other charged particles in secondary cosmic
ray flux) is just the same order of magnitude\cite{Shen} as Luo's
obtained upper limit on photon mass.

It is known that at the sea level, the current density of muon
component in secondary cosmic rays is about $1\times 10^{-2}{\rm
cm}^{-2}\cdot{\rm s}^{-1}$. Assuming that the muon velocity
approaches speed of light, the volume density of muon can be
derived and the result is $N_{\mu}=0.3\times 10^{-6}$ ${\rm
m}^{-3}$. So, according to the {\it photon effective rest mass}
formula $m_{\rm
eff}=\frac{\hbar}{c}\sqrt{\frac{2N_{\mu}e^{2}}{\epsilon_{0}m_{\mu}c^{2}}}$\cite{Shen}
with $m_{\mu}$, $\hbar$, $c$ and $\epsilon_{0}$ denoting the muon
mass, Planck's constant, speed of light and electric permittivity
in a vacuum, respectively, electromagnetic wave with wavelength
$\lambda \gg 100$ m at the sea level acquires an effective rest
mass about $0.3\times 10^{-53}$ Kg. Hence the ambient cosmic
magnetic vector potentials (interstellar magnetic fields) with low
or zero frequencies will truly acquire this effective rest mass.

Note that the above-mentioned muon volume density ($0.3\times
10^{-6}$ ${\rm m}^{-3}$) is only the datum at the sea level. In
order to evaluate the muon volume density in the environment where
Luo's experiment was performed, readers may be referred to the
following handbook data of muon current density in underground
cosmic rays: the muon current densities are respectively $10^{-4}$
${\rm cm}^{-2}\cdot{\rm s}^{-1}$ and $10^{-6}$ ${\rm
cm}^{-2}\cdot{\rm s}^{-1}$ at the equivalent water depths of $100$
m and $1000$ m under the ground. In Luo's experiment, the total
apparatus is located in a cave laboratory, on which the least
thickness of the cover is more than $40$ m\cite{Luo}. It is
reasonably believed that the muon density of secondary cosmic rays
in the vacuum chamber of Luo's experiment may be one or two orders
of magnitude less than that at the sea level. This, therefore,
means that the effective rest mass acquired by photons in Luo's
vacuum chamber may be about $(0.3\sim 1)\times 10^{-54}$ Kg.

It is believed that the air molecules in the low-pressure vacuum
chamber of torsion balance experiment cannot be easily ionized by
the alpha-particles, electrons and protons of cosmic rays,
because, for example, it is readily verified that the mean free
path of alpha-particle moving at about $10^{7}{\rm m}\cdot{\rm
s}^{-1}$ in the dilute air with the pressure being only $10^{-2}$
Pa\cite{Luo} is too large ({\it i.e.}, more than $10^{6}$
m)\cite{Shen}. So, the air medium in the low-pressure vacuum
chamber has nothing to contribute to the effective rest mass of
photons. However, as far as the alpha-particles, $\pi$ mesons,
electrons, {\it etc.} in secondary cosmic rays are concerned, it
follows that these ions will also contribute $10^{-54}$ Kg or so
to photon effective mass, since their total volume density in
Luo's cave laboratory is about only one order of magnitude less
than that of muon.

Thus the photon effective rest mass resulting from both muons and
other ions in secondary cosmic rays can be compared to the newly
obtained upper limit ($1.2\times 10^{-54}$ Kg) by Luo {\it et al.}
in their recent rotating torsion balance experiment\cite{Luo}. For
this reason, now Luo's experimental upper limit on photon mass
becomes just an interesting critical value. We think Luo's recent
result is of physical interest but it still deserves further
experimental investigation so as to improve the present upper
limit on photon mass. It is claimed that, in the near future, only
the contribution of ions in secondary cosmic rays is ruled out,
can we deal better with the experiment schemes of photon rest mass
and related experimental results.

\textbf{Acknowledgements}  This project is supported partially by
the National Natural Science Foundation of China under the project
No. $90101024$.

\end{document}